\documentclass[conference]{IEEEtran}
\usepackage{graphicx}
\usepackage{caption}
\usepackage{mwe}
\usepackage{subfig}
\usepackage{amsmath,amsfonts,amssymb}
\usepackage{bbm}

\usepackage{color}
\usepackage{xcolor}
\usepackage[linesnumbered, ruled,noend]{algorithm2e}
\makeatletter
\renewcommand{\@algocf@capt@plain}{above}
\makeatother

\SetCommentSty{mycommfont}
\DeclareMathOperator*{\argmax}{arg\,max}
\usepackage{makecell}

\newsubfloat{figure}

\usepackage{nth}
\usepackage{mdwtab}
\usepackage{enumitem}
\usepackage{bm}

\usepackage{cite}

\ifCLASSINFOpdf
\else
\fi
\usepackage{url}

\renewcommand{\baselinestretch}{0.99}

\begin{document}
%
\title{CRAD: Clustering with Robust Autocuts and Depth}

\author{\IEEEauthorblockN{Xin Huang}
\IEEEauthorblockA{Department of Mathematical Sciences \\
University of Texas at Dallas\\
Richardson, USA\\
Email: xxh130130@utdallas.edu}
\and
\IEEEauthorblockN{Yulia R. Gel}
\IEEEauthorblockA{Department of Mathematical Sciences\\
University of Texas at Dallas\\
Richardson, USA\\
Email: ygl@utdallas.edu}
}


%


\maketitle

\begin{abstract}
We develop a new density-based clustering algorithm named CRAD which is based on a new neighbor searching function with a robust data depth as the dissimilarity measure. Our experiments prove that the new CRAD is highly competitive at detecting clusters with varying densities, compared with the existing algorithms such as DBSCAN, OPTICS and DBCA. 
Furthermore, a new effective parameter selection procedure is developed to select the optimal underlying parameter in the real-world clustering, when the ground truth is unknown. Lastly, we suggest a new clustering framework that extends CRAD from spatial data clustering to time series clustering without a-priori knowledge of the true number of clusters. The performance of CRAD is evaluated through extensive experimental studies.
\end{abstract}

\textbf{\textit{keywords-clustering}}, \textbf{\textit{space-time processes}}, \textbf{\textit{data depth}}

\ifCLASSOPTIONpeerreview
\begin{center} \bfseries EDICS Category: 3-BBND \end{center}
\fi
%
\IEEEpeerreviewmaketitle

\section{Introduction}\label{section:introduction}

Data depth methodology is a widely employed nonparametric tool in multivariate and functional data analysis, with applications ranging from outlier detection to clustering and visualization~\cite{
liu1999multivariate,
cuevas2007robust,li2012dd}. Depth measures the ``centrality'' (or ``outlyingness'') of a given object 
with respect to an observed data cloud~\cite{aMosler2012,zuo2000general}. Many desirable properties of data depth such as affine invariance, robustness, and center maximality have earned it an increasing attention in the machine learning and statistics communities in the last decade. There exist numerous clustering and classification methods, based on a data depth concept~\cite{
jornsten2004clustering, aMosler2012, pokotylo2016depth}. Most such methods, however, rely on the knowledge of a true number of clusters $k$.
Most recently,~\cite{jeong2016data} proposed a Depth Based Clustering Algorithm (DBCA) and showed benefits of a data depth for clustering spatial data. 
To the best of our knowledge,~\cite{jeong2016data} is the first and only reference introducing a data depth concept into clustering analysis of spatial data. However, DBCA cannot handle a case of clusters with varying densities, which is a common issue in density-based clustering domain. In addition, the problem how to select the tuning parameter, which highly impacts the clustering result, remains unstudied.

The current paper is motivated by three over-arching major challenges in density-based clustering: (1) Based on data depth, can we propose an algorithm that delivers more robust performance under the existence of clusters with varying densities? (2) Based on the proposed algorithm, how can we select the true underlying parameter in the real-world clustering when the ground truth is not given? (3) Can the density-based algorithm be extended to multivariate time-series clustering, without a-priori knowledge of the number of clusters? We address these three major problems by proposing a new clustering algorithm, named Clustering with Robust Autocuts and Depth (CRAD).



\begin{figure*}
\minipage{0.5\textwidth}
\centering
	\includegraphics[width=\linewidth]{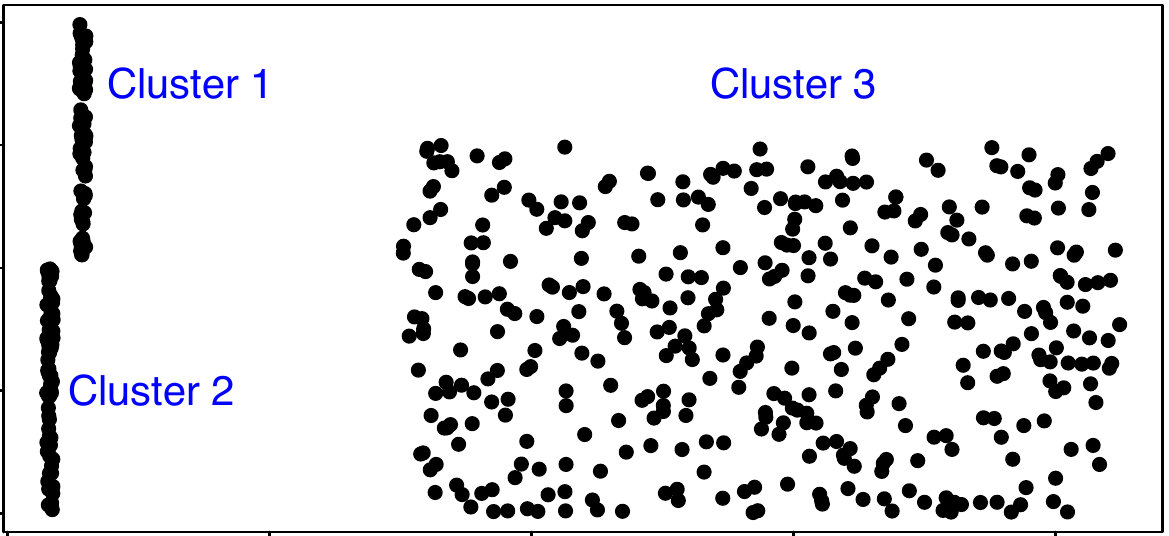}
	\caption*{\small(a) Raw Data}
\endminipage\hspace{0.2cm}%
\minipage{0.5\textwidth}
\centering
  \includegraphics[width=\linewidth]{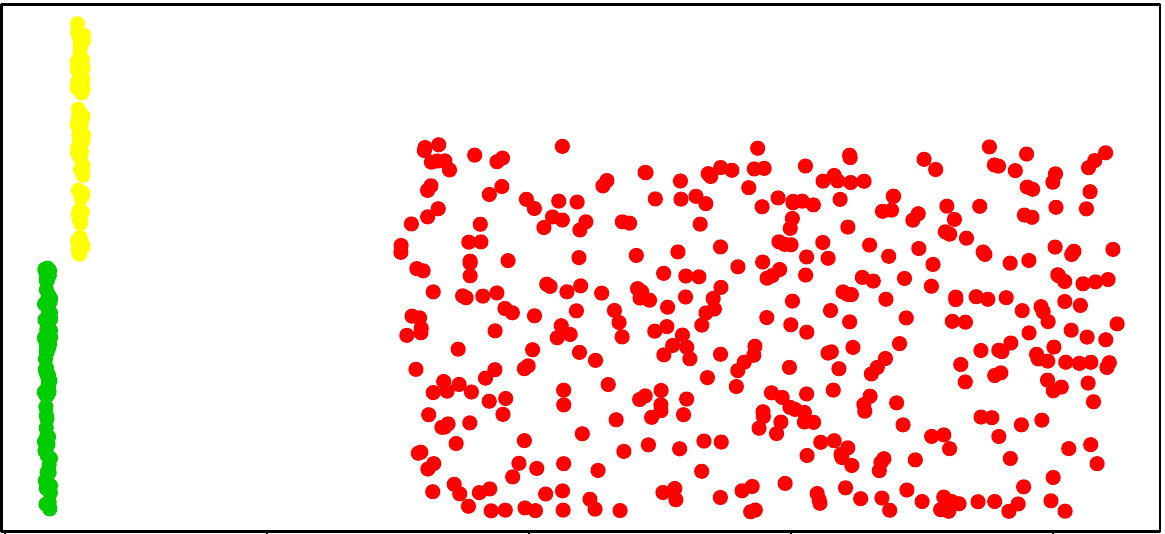}
  \caption*{\small(b) CRAD}
\endminipage
\vspace*{0.3cm}

\minipage{0.5\textwidth}%
\centering
  \includegraphics[width=\linewidth]{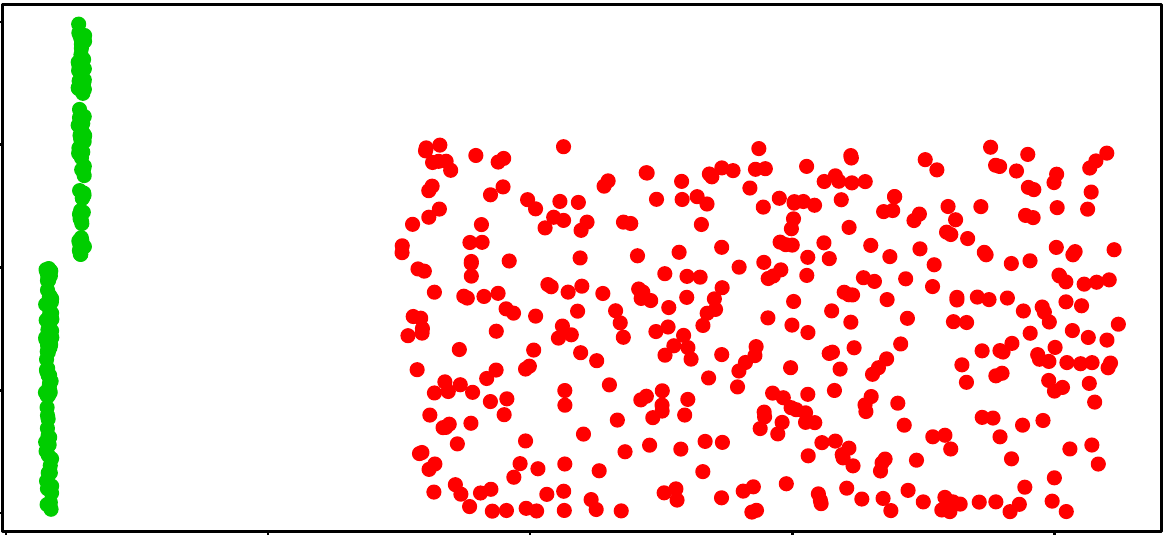}
  \caption*{\small(c) DBCA and DBSCAN}
\endminipage\hspace{0.2cm}%
\minipage{0.5\textwidth}%
\centering
  \includegraphics[width=\linewidth]{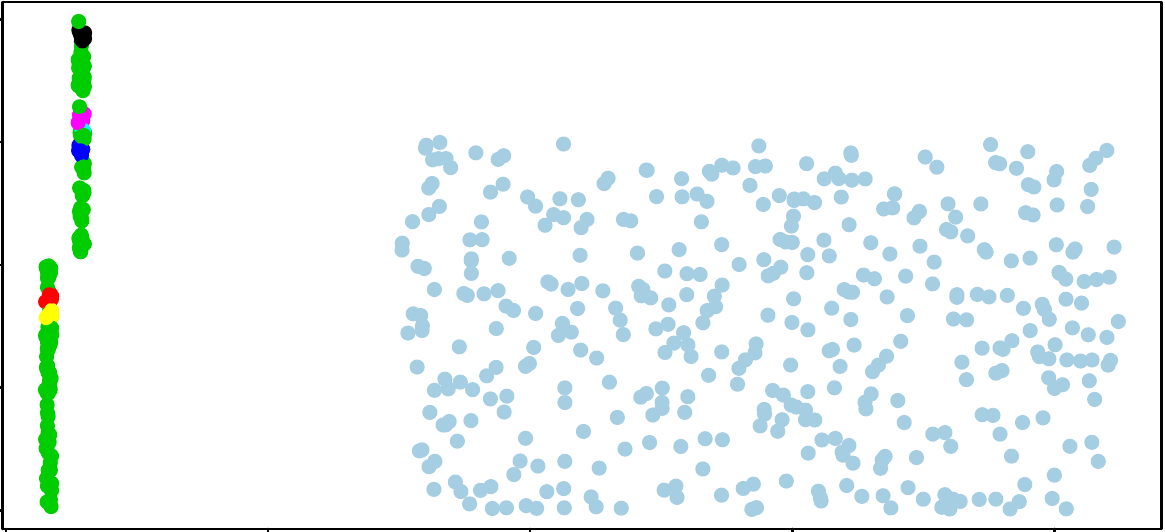}
  \caption*{\small(d) OPTICS}
\endminipage

\caption{Clustering Performance of CRAD, DBCA, DBSCAN and OPTICS on the Toy Example.}\label{fig: toy ex}
\end{figure*}

One of the key benefits of the new CRAD algorithm is its ability to detect clusters with varying densities. Let us start with a simple yet typical dataset to shed some light on the difference between our algorithm and some existing algorithms such as DBSCAN~\cite{ester1996density}, OPTICS~\cite{ankerst1999optics}, and DBCA~\cite{jeong2016data} in addressing this type of problem. As shown in Fig.~\ref{fig: toy ex}(a), the toy dataset includes two dense clusters (clusters 1 and 2), and one sparse cluster (cluster 3). The number of observations in cluster 3 is larger than that in clusters 1 and 2. The result of each algorithm is selected by searching the best clustering performance on a wide range of possible combinations of its tuning parameters.
Clustering results are shown in Fig.~\ref{fig: toy ex}. Currently available methods such as DBCA, DBSCAN, and OPTICS, all fail to separate the cluster 1 and 2; in contrast, our new CRAD algorithm is able to detect both. The reason for this phenomenon is that both DBSCAN and DBCA use  globally-defined parameters (i.e., $\epsilon$ and $\theta$, respectively) to find clusters, thus lacking the flexibility to adjust their value when clusters have different densities. Even OPTICS, which is proposed to solve this density variation problem, still does not deliver competitive clustering performance on the toy example. Our algorithm, in contrast, uses a \textit{locally-defined} parameter to customize the neighbor searching function for each observation, based on a notion of density level. As a result, CRAD is able to deliver a competitive performance in separating clusters with varying densities. 

This paper makes the following novel contributions to spatial and temporal clustering:
\begin{enumerate}[leftmargin=*]
 \item We propose a new robust density-based clustering algorithm (CRAD), using a notion of statistical \textit{data depth} as the dissimilarity measure, and further augment the depth-based clustering analysis with an outlier-resistant and highly computationally efficient estimator of multivariate scale, namely, the Minimum Covariance Determinant (MCD).
      Our experiments prove that the new algorithm CRAD is highly competitive at detecting clusters with varying densities, compared with the existing algorithms such as DBSCAN, OPTICS and DBCA.
 \item   Furthermore, we show that a hybrid combination of our new robust depth-based neighbor searching algorithm and conventional DBSCAN, allows to significantly improve clustering performance of DBSCAN. This is an important standalone step toward future extension of DBSCAN to non-Euclidian spaces and functional data clustering.
 \item We develop a new effective parameter selection procedure to select the optimal underlying parameter in the real-world clustering, when the ground truth is unknown.
  \item We suggest a new clustering framework that extends CRAD from spatial data clustering to time series clustering without a-priori knowledge of the true number of clusters. Performance of CRAD in time series clustering is evaluated with extensive experiments on benchmark data.
\end{enumerate}

%

	The paper is organized as follows. In Section~\ref{section:algo} we present the new algorithm CRAD. In Section~\ref{section: determine para} an effective parameter selection procedure is proposed to select the parameters in CRAD. We evaluate CRAD through extensive numerical studies in Section~\ref{section: experimental evaluation}. The paper is concluded with discussion and future research directions in Section~\ref{section: conclusion}.


\section{Our Algorithm}\label{section:algo}

We start from providing a direct insight into our algorithm, with a particular emphasis on introducing the distinguishing features of CRAD, namely, the dissimilarity measure and the neighbor searching function.

Before proceeding to details, we first review the general structure for density-based clustering algorithm. Let the data be stored as an $n\times p$-matrix $\mathbf{X} = (\mathbf{x}_1,\dots ,\mathbf{x}_n )^t$ with $\mathbf{x}_i = (x_{i1},\dots, x_{ip})^t$ the $i$-th observation, and $n$ be a sample size. The core idea behind all density-based algorithms is to assign a $\{0,1\}$-relationship between all observations in $\mathbf{X}$, based on how close the two observations are, in terms of a given dissimilarity measure. That is, a neighbor searching function $\text{NBR}(\mathbf{x}_i), i = 1,2,\dots,n$ is needed such that $\mathbf{x}_i$ and $\mathbf{x}_j$ are $1$-related if $\mathbf{x}_j\in \text{NBR}(\mathbf{x}_i)$ and $0$-non-related if $\mathbf{x}_j\notin \text{NBR}(\mathbf{x}_i)$. These results are stored in a $\{0,1\}$-adjacency matrix $A$, and
a breadth-first search is then applied to $A$ to generate the final clustering partition of $\mathbf{X}$. What discriminates the clustering algorithms, however, is the \textit{dissimilarity measure} and \textit{neighbor searching function}. We present the distinguishing features of CRAD in terms of these two as follows.

\subsection{A Robust Data Depth Based Dissimilarity}
A data depth is a function that quantifies how closely
an observed point $x \in \mathbb{R}^d$, $d\geq 2$, is located to the ``center'' of a finite set $\mathcal{X}\in \mathbb{R}^d$, or relative to a probability distribution $P$ in $\mathbb{R}^d$. A data depth shall satisfy the following desirable properties (\cite{aMosler2012, zuo2000general}): affine invariant; upper semi-continuous in $x$; quasiconcave in $x$; (i.e., having convex upper level sets) vanishing as $||x||\to \infty$ (\cite{aMosler2012, zuo2000general}).

We propose to utilize a robust Mahalanobis depth function, with the Minimum Covariance Determinant (MCD) as an outlier-resistant and highly computationally efficient estimator of multivariate scale, as an alternative clustering dissimilarity measure. That is, let the data be stored as an $n\times p$-matrix $\mathbf{X} = (\mathbf{x}_1,\dots ,\mathbf{x}_n )^t$ with $\mathbf{x}_i = (x_{i1},\dots, x_{ip})^t$ the $i$-th observation, and $n$ be a sample size. The Robust Mahalanobis depth function can be defined as:
\begin{equation}\label{mahala}
	RM_d(\mathbf{x}_j|\mathbf{x}_i) = [1 + (\mathbf{x}_j - \mathbf{x}_i)^T \mathbf{\Sigma}^{-1}(\mathbf{x}_j - \mathbf{x}_i)]^{-1},
\end{equation}
where 
$$\mathbf{\Sigma} = c_1 \frac{1}{n} \sum\limits_{i=1}^{n} W(d_i^2)(\mathbf{x}_i - \hat{\mathbf{u}}_{MCD})(\mathbf{x}_i - \hat{\mathbf{u}}_{MCD})^T,$$
and $\hat{\mathbf{u}}_{MCD} = \sum_{i=1}^n W(d_i^2)\mathbf{x}_i/ \sum_{i=1}^n W(d_i^2)$; $ d_i = \sqrt{(\mathbf{x}_i - \mathbf{\hat{\mu}}_o)^T \mathbf{\hat{\Sigma}_o}^{-1}(\mathbf{x}_i - \mathbf{\hat{\mu}}_o)}$; $W$ is an appropriate weight function; $\hat{\mathbf{\mu}}_0$ and $\mathbf{\hat{\Sigma}}_o$ are sample mean and sample covariance matrix, respectively; and $c_1$ is a consistency factor~\cite{hubert2010minimum}. The MCD covariance estimator has been proven to significantly outperform the Minimum Volume Ellipsoid (MVE) covariance estimator,
that is used by~\cite{jeong2016data}, both in terms of statistical efficiency and computation (see, e.g.,~\cite{van2009minimum}). 
The high computational efficiency of MCD makes it a preferred method over MVE, especially in modern high dimensional problems.

\begin{figure}[!htb]

\centering
	\includegraphics[width=\linewidth]{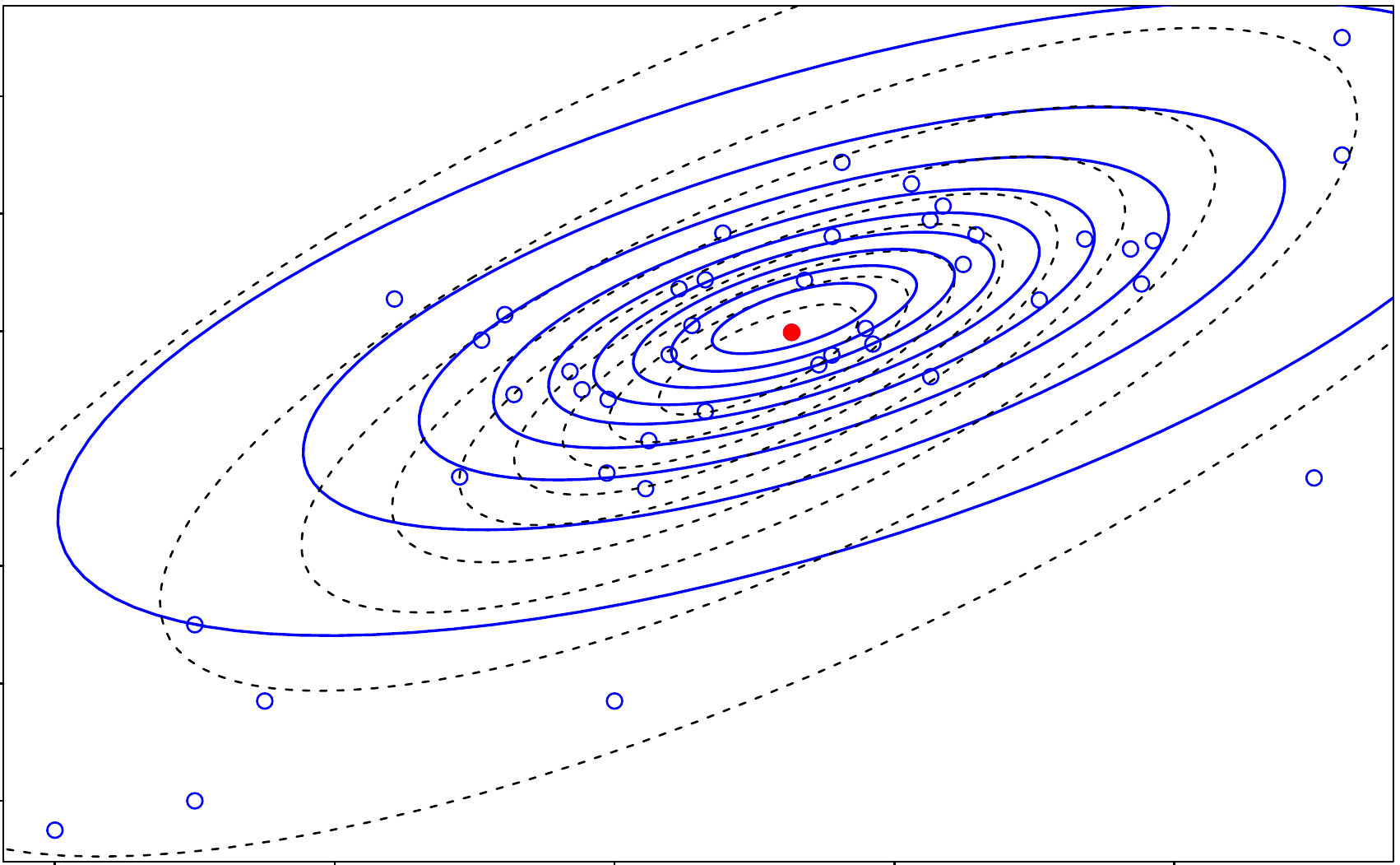}

\caption{A contour plot based on traditional (black dash line) and robust (blue solid line) Mahalanobis depth function.}
  \label{fig:Contour Plot}
\end{figure}

Now for each $\mathbf{x}_i$, we calculate a robust Mahalanobis depth vector $\mathbf{RM}_{d}(\mathbf{x}_i)=\langle RM_d(\mathbf{x}_1|\mathbf{x}_i), \ldots,RM_d(\mathbf{x}_{n}|\mathbf{x}_i)\rangle$, measuring the ``outlyingness" of every other observation with respect to $\mathbf{x}_i, i=1,2,\dots,n$. The depth vector $\mathbf{RM}_{d}(\mathbf{x}_i)$
provides a center-outward ordering of the data and serves as a topological map. The effect of traditional and robust Mahalanobis depth function is visualized in Fig.~\ref{fig:Contour Plot}, where the solid red dot represents the observation $\mathbf{x}_i$ (center) and each contour corresponds to a depth value. Armed with a robust depth-based dissimilarity measure~(\ref{mahala}), we now proceed to clustering. 

\subsection{A New Neighbor Searching Algorithm}

\subsubsection{Who is Your Closest Neighbor?}
In CRAD, we use a robust depth-based dissimilarity measure~(\ref{mahala}), and the neighbor searching function is defined as:
\begin{equation}\label{equ:crad}
\begin{split}
\hspace*{-1.3mm}\text{NBR}(\mathbf{x}_i) = \{\mathbf{x}_j\colon RM_d(\mathbf{x}_j|\mathbf{x}_i) \geq h_{opt}(i), j = 1,\ldots,n \},
\end{split}
\end{equation}
where $RM_d(\mathbf{x}_j|\mathbf{x}_i)$ is defined in~(\ref{mahala}) and $h_{opt}(i)$ is the cut-off parameter. The novel part of our neighbor searching function is that for each observation $\mathbf{x}_i, i=1,\ldots, n$, the cut-off parameter $h_{opt}(i)$ is \textit{locally} rather than globally defined, and accounts for different density level around it. E.g., if a person resides in Manhattan, his closest neighbor is likely in the same apartment complex; but if he lives in Dallas, TX, the closest neighbor might be miles away.



In contrast, DBCA uses a globally defined parameter $\theta$ in its neighbor searching function:
\begin{equation}\label{equ:dbca}
\begin{split}
\text{NBR}(\mathbf{x}_i) =\{\mathbf{x}_j\colon RM_d(\mathbf{x}_j|\mathbf{x}_i) \geq \theta, j = 1,\ldots,n \},
\end{split}
\end{equation}
Similarly, DBSCAN uses a globally-defined parameter $\epsilon$ in its neighbor searching function:
\begin{equation}\label{equ:dbscan}
\begin{split}
\text{NBR}(\mathbf{x}_i) =\{\mathbf{x}_j\colon \lVert \mathbf{x}_i - \mathbf{x}_j \rVert_2 \leq \epsilon, j = 1,\ldots,n \}.
\end{split}	
\end{equation}
With an additional requirement on the minimum number of observations $MinPts$ in each cluster, elements $A_{ij}$ of the adjacency matrix for DBSCAN are defined as 1 if $\mathbf{x}_j \in \text{NBR}(\mathbf{x}_i)$ and $|\text{NBR}(\mathbf{x}_i)| > MinPts$, and 0, otherwise.
%
Here $|X|$ denotes the cardinality of a set $X$.

\begin{figure*}
\centering
  \includegraphics[scale=0.50]{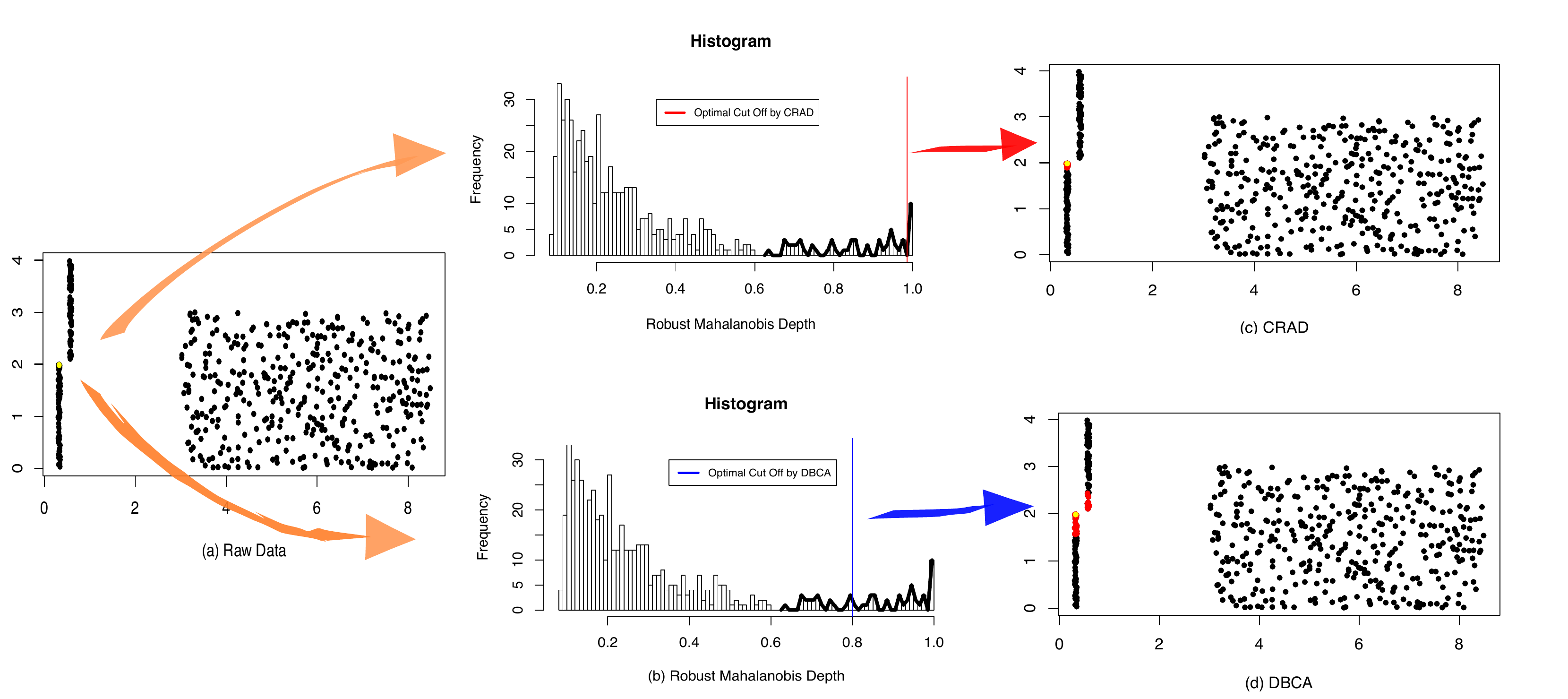}\caption{Neighbor search for a given point (the yellow dot) in the toy example. The red dots are the neighbors, identified by CRAD and DBCA under their best clustering performances, shown in top (bottom) right. The histogram shows the optimal value of the cut-off parameter, $h_{opt}(i)$ (the red vertical line) for CRAD and $\theta$ (the blue vertical line) for DBCA.}\label{fig:Big}
\vspace*{-0.5cm}
\end{figure*}

The parameter in $\text{NBR}$ is critical in detecting cluster patterns. A globally-defined parameter cannot find all intrinsic clusters with varying densities. The example in Fig.~\ref{fig:Big} best illustrates the idea: we take the toy data in Section~\ref{section:introduction} and investigate the neighbor searching process of an observation for CRAD and DBCA~\cite{jeong2016data}. The observation is labeled as the yellow dot in Fig.~\ref{fig:Big}(a). The reader could visualize the difference between the neighbor observations (red dots) found by the locally-adjusted parameter $h_{opt}(i)$ in CRAD and the globally-defined parameter $\theta$ in DBCA, as shown in the Fig.~\ref{fig:Big}(c), (d). We find that the DBCA incorrectly includes the observations in the nearby cluster of the yellow dot as its neighbors, thus leading to the inaccurate clustering result. 
Given the ground truth, the parameter $\theta$ in DBCA is selected by searching the best clustering result over a wide range of values $[0.80, 0.81,\ldots,1]$. In contrast, parameter $h_{opt}(i)$ of CRAD is selected by an automatic self-searching algorithm based on a notion of density level (see Algorithm~\ref{algo: neighbor function}). 

\subsubsection{An Automatic Self-Searching Algorithm for Finding $h_{opt}(i)$}\label{subsubsection: self searching}
The idea is that the neighbor searching function of each observation should depend on the \emph{relative change of the density level} around it. The term ``relative" accounts for the customization for each observation.
As mentioned before, for each $\mathbf{x}_i$ we calculate a robust Mahalanobis depth vector
$\mathbf{RM}_{d}(\mathbf{x}_i)=\langle RM_d(\mathbf{x}_1|\mathbf{x}_i), \ldots,RM_d(\mathbf{x}_{n}|\mathbf{x}_i)\rangle$, measuring the ``outlyingness" of every other observation with respect to $\mathbf{x}_i, i=1,2,\dots,n$ (Fig.~\ref{fig:Contour Plot}). 



Armed with $\mathbf{RM}_{d}(\mathbf{x}_i)$ of $\mathbf{x}_i$, we create a vector of histogram
$\mathbf{H} = \langle h_{width}, h_{2*width}, \ldots, h_1 \rangle$,
where $h_j = \sum_{k=1}^n \mathbbm{1}_{j - width} <RM_{d}(\mathbf{x}_k|\mathbf{x}_i) \leq j $. Parameter $width = 1 / Nbin \in (0,1)$, where $Nbin$ is the number of bins in $\mathbf{H}$ and is user pre-defined. Analogous to the definition of density for a substance, $\rho = m/V$ mass ($m$) per unit volume ($V$), we define the density level of a point as $N/d$, number of observations ($N$) per unit depth distance ($d$). If we choose the unit depth distance as the parameter $width$,  then the reverse order of $\mathbf{H}$: $h_1, h_{1-width},h_{1-2*width},\ldots, h_{width}$ are the density levels around $x_i$ in a center-outward order. A higher value of $h_k, k= 1,1-width,1-2*width,\dots,width$ indicates a denser region and a lower value corresponds to a sparser region. Thus, starting from $h_1$ we search for the first local minimum $h_{opt}$ over $\mathbf{H}$. The value of $h_k, k=1,1-width,1-2*width,\ldots,width$ decreasing from $h_1$ to $h_{opt}$ indicates that the density level around $\mathbf{x}_i$, in a center-outward manner, changes from dense to sparse. The observations in the sparse region do not have the same property as the observations in the dense region. Thus, the first local minimum $h_{opt}$ could serve as the cut-off depth value to select the neighbors of $\mathbf{x}_i$. For each $\mathbf{x}_i,i=1,2,\dots,n$, a locally-defined $h_{opt}(i)$ is selected. Thus, neighbor observations of $\mathbf{x}_i$ can be found from~(\ref{equ:crad}).
Fig.~\ref{fig:Big}(b) shows how the neighbor searching parameter $h_{opt}(i)$ (red vertical line) is selected for each $\mathbf{x}_i$. Note, the DBCA does not include a similar self-searching step. For better comparison and visualization purpose we put the selected $\theta$ (blue line) in the histogram plot.

The CRAD algorithm is summarized in Algorithm~\ref{CRAD algorithm}. The neighbor searching function and the automatic self-searching method are described in Algorithm~\ref{algo: neighbor function}. 
A user pre-defined parameter $StepSize$ is required to decide the size of neighbor buckets in $\mathbf{H}$ to compare for each $h_i, i = 1, 1-width,1-2*width,\ldots,width$. Another user pre-defined parameter is the number of bins $Nbin$ in generating $\mathbf{H}$ (for details on $Nbin$ selection see Section~\ref{section: determine para}). The upper bound for time complexity of CRAD is $O(n^2)$, and complexity can be further lowered to achieve $O(n\log{}n)$, by using an accelerating index structure for the data in two dimensional spaces~\cite{ester1996density, gan2015dbscan}. The source code of CRAD algorithm is available from {\small \url{https://github.com/DataMining-ClusteringAnalysis/CRAD-Clustering/}}.

%

{\begin{algorithm}
\KwIn{A finite set of observations $\mathbf{X} = (\mathbf{x}_1,\dots ,\mathbf{x}_n )^t$ with $\mathbf{x}_i = (x_{i1},\dots, x_{ip})^t$ the $i$th observation; $n$: Sample size; $Nbin$: Number of bins; $StepSize$: Size of neighbor buckets to compare.}
\KwOut{$\text{ClVec}$: Cluster ID of each observation. }
Initialization: $\text{ClVec} = [-1]*n$; $\text{label} := 0$; $A := \{0\}^{n\times n}$\; 
Compute Robust Mahalanobis depth vector for each observation: $\mathbf{RM}_{d} = \langle \mathbf{RM}_{d}(\mathbf{x}_1), \mathbf{RM}_{d}(\mathbf{x}_2), \dots , \mathbf{RM}_{d}(\mathbf{x}_n) \rangle$\;
\tcp{Compute the adjacency matrix $A$}
\For{$i := 1$ \textbf{to} $n$}{
	
	$\text{AdjIndex} := \text{NBR}(\mathbf{RM}_d(\mathbf{x}_i), Nbin, StepSize)$;\
	$\text{for}\ \forall \ \text{ind} \in \text{AdjIndex}: A[i, \text{ind}] := 1$;\

}

\For{$i := 1$ \textbf{to} $n$} {
  \If{$\text{ClVec} [i] == -1$} {
    $\text{nbrs}:= \text{neighbor IDs of observation i}$\;

    \If{$\text{nbrs. size} () == 1$}{
    	$\text{ClVec}[i] := 0$; \tcp{single cluster}
    	}
    \Else{
    	$\text{label} := \text{label} + 1$\;
    	$\text{for}\ \forall \ \text{nbrId} \in \text{nbrs}: \text{ClVec}[\text{nbrId}] := \text{label} $\;
    	$\text{nbrs. remove}(i)$\;
    	
    	\While{$\text{nbrs is not empty}$}{
    	
    		$\text{CurrentPoint} := \text{nbrs. get}()$
    	
    		$\text{Snbrs} := \text{neighbor IDs of CurrentPoint}$;
    		
    		\If{$\text{Snbrs. size}() > 1$}{
    			
    			\For{$x \ \text{in Snbrs}$}{
    			
    				\If{$\text{ClVec}[x] == -1$}{
    					$\text{ClVec}[x] := \text{label}$;
    					$\text{nbrs. add}(x)$;\				
    				}
    			}
    		}
    		
    	}
   }

  }
}
\caption{CRAD Algorithm}\label{CRAD algorithm}
\end{algorithm}
}
%
%
%
%
%
%

{
\begin{algorithm}
\KwIn{$\mathbf{RM}_d(\mathbf{x}_i)$: Robust Mahalanobis depth vector of observation $i$; $Nbin$: Number of bins; $StepSize$: Size of neighbor buckets to compare.}
\KwOut{$\text{nbrIds}$: Neighbor IDs of observation $i$.}
Initialization: $width := 1/Nbin$.

Compute histogram $H$ based on $\mathbf{RM}_d(\mathbf{x}_i)$: $ H = \langle h_{width}, h_{2*width}, \ldots, h_1 \rangle$.

\caption{  $\text{NBR} (\mathbf{RM}_d(\mathbf{x}_i),Nbin,StepSize)$ } \label{algo: neighbor function}

\For{$j := H. size() - StepSize$ \textbf{to} $1 + StepSize$}{

	$\text{Boolean b}:= \text{An empty array}$;\
	
	\For{$z:= 1$ \textbf{to} $StepSize$}{
		
		\If{$H[j] < H[j+z]$ \textbf{and} $H[j] < H[j-z]$}{
		
			$b.append (\text{TRUE})$;\
		}
		\Else{$b.append (\text{FALSE})$;\
		}	
	}
	
	\If{$(b == \text{TRUE}).size() == StepSize$}{
		
		$h_{opt} := 1 - (H. size()-j+1)*width$;\
	
		\textbf{Break};
	}
}

$\text{nbrIds} := \{l: RM_d(\mathbf{x}_l|\mathbf{x}_i) > h_{opt}, l = 1,\dots,n\}$.

\end{algorithm}
}
%

\subsection{An Extension to DBSCAN}\label{subsec:extension to dbscan}
Since the essential difference between CRAD and DBSCAN is the neighbor searching function, a hybrid combination of our new robust depth-based neighbor searching algorithm and conventional DBSCAN is generated by replacing the neighbor searching function~(\ref{equ:dbscan}) in DBSCAN with the proposed new function~(\ref{equ:crad}). We name the hybrid algorithm as CRAD-DBSCAN. Our experiments show that with a replacement of a neighbor searching function, CRAD-DBSCAN significantly outperforms DBSCAN (see Section~\ref{subsec:synthetic data}).
This is an important standalone step toward future extension of DBSCAN to non-Euclidian spaces and functional data clustering. That is, the DBSCAN approach and its adaptations, such as CRAD-DBSCAN, with a suitable metric as a dissimilarity measure (e.g., band depth), can be further advanced to clustering of functional curves in Hilbert spaces.

\section{Determining the Parameter $StepSize$ and  $Nbin$}\label{section: determine para}

Our CRAD algorithm requires two parameters, $StepSize$ and $Nbin$, both of which are used in the automatic self-searching algorithm in Algorithm~\ref{algo: neighbor function}. The goal is to select optimal $StepSize$ and $Nbin$ to help CRAD achieve the highest quality of clustering results.
There are two kinds of evaluation metrics to measure the quality of clustering results, external and internal metrics. An external metric, such as Rand Index (RI)~\cite{rand1971objective, jain1999data} and Adjusted Mutual Information (AMI)~\cite{
meilua2007comparing}, is a measure of agreement between the result obtained from a clustering algorithm and the  ground truth.
Since the ground truth is not available in the real-world clustering, we use the internal metric, which measures the goodness of clustering without external information~\cite{rousseeuw1987silhouettes, calinski1974dendrite, de2015recovering}, to serve as a validation tool for selecting optimal $StepSize$ and $Nbin$. If we assume larger values of the metric indicate better clustering results, $StepSize_{opt}$ and $Nbin_{opt}$ are then
defined as:
\begin{equation}\label{equ: parameter decision}
\begin{split}
StepSize_{opt}, Nbin_{opt} :=\argmax\limits_{StepSize,Nbin} M (\mathbf{X}, \text{ClVec}),
\end{split}
\end{equation}
where {$\text{ClVec}$} is the clustering result returned by Algorithm~\ref{CRAD algorithm}. 
Here we consider the Calinski-Harabasz (CH) score as the internal metric $M$, which evaluates the clustering quality based on the average between- and within-cluster sum of squares~\cite{calinski1974dendrite, frakes1992information}.

The CH score is defined as:
\begin{equation}
	CH(\mathbf{X}, ClVec) = \frac{trace \mathbf{B}/(k - 1)}{trace \mathbf{W}/(n - k)},
\end{equation}
where $\mathbf{B}$ is the error sum of squares between different clusters (between-cluster),
\begin{equation}
	trace \mathbf{B} = \sum\limits_{m=1}^k |\overline{C}_m|\lVert \overline{C}_m - \bar{\mathbf{x}} \rVert_2,
\end{equation}
and $\mathbf{W}$ is the squared differences of all objects in a cluster from their respective cluster center (within-cluster)
\begin{equation}
    trace  \mathbf{W} = \sum\limits_{m=1}^k\sum\limits_{i=1}^{n} w_{m,i} \lVert \mathbf{x}_m- \overline{C}_m\rVert.
\end{equation}
Here $|\bar{C}_m|$ and $\bar{\mathbf{x}}$ are the sample mean of $m$th cluster and the data set $\mathbf{X}$, respectively; $n$ is sample size; $k$ is the number of clusters in $ClVec$, and $w_{m,i}$ is the weight function. The larger value of $CH$, the better clustering performance~\cite{calinski1974dendrite}.

Our simulations show that optimal performance can be achieved with $StepSize \in \{1, 2\}$  and $Nbin \in (0.2*n-100, 0.2*n+100)$, where $n$ is the sample size. Thus, we fix $StepSize$ as 1 and search $Nbin_{opt}$ based on~(\ref{equ: parameter decision}). 
(for details see Section~\ref{section: experimental evaluation}).


\section{Experimental Evaluation}\label{section: experimental evaluation}

\subsection{Synthetic Data}\label{subsec:synthetic data}

We evaluate performance of CRAD with respect to DBCA~\cite{jeong2016data}, DBSCAN~\cite{ester1996density}, and OPTICS~\cite{ankerst1999optics}. The DBSCAN has two versions: 1. Original DBSCAN with Euclidean distance as the dissimilarity measure, i.e., DBSCAN (EU); 2. An extension version of DBSCAN (CRAD-DBSCAN), which is a hybrid combination of our new robust depth-based neighbor searching algorithm and conventional DBSCAN, as discussed in Section~\ref{subsec:extension to dbscan}. We show that with a simple replacement of a neighbor searching function, CRAD-DBSCAN significantly outperforms DBSCAN in the considered set of synthetic data.

\begin{figure*}
\centering
\minipage{0.3\textwidth}
\centering
  \includegraphics[width=\linewidth]{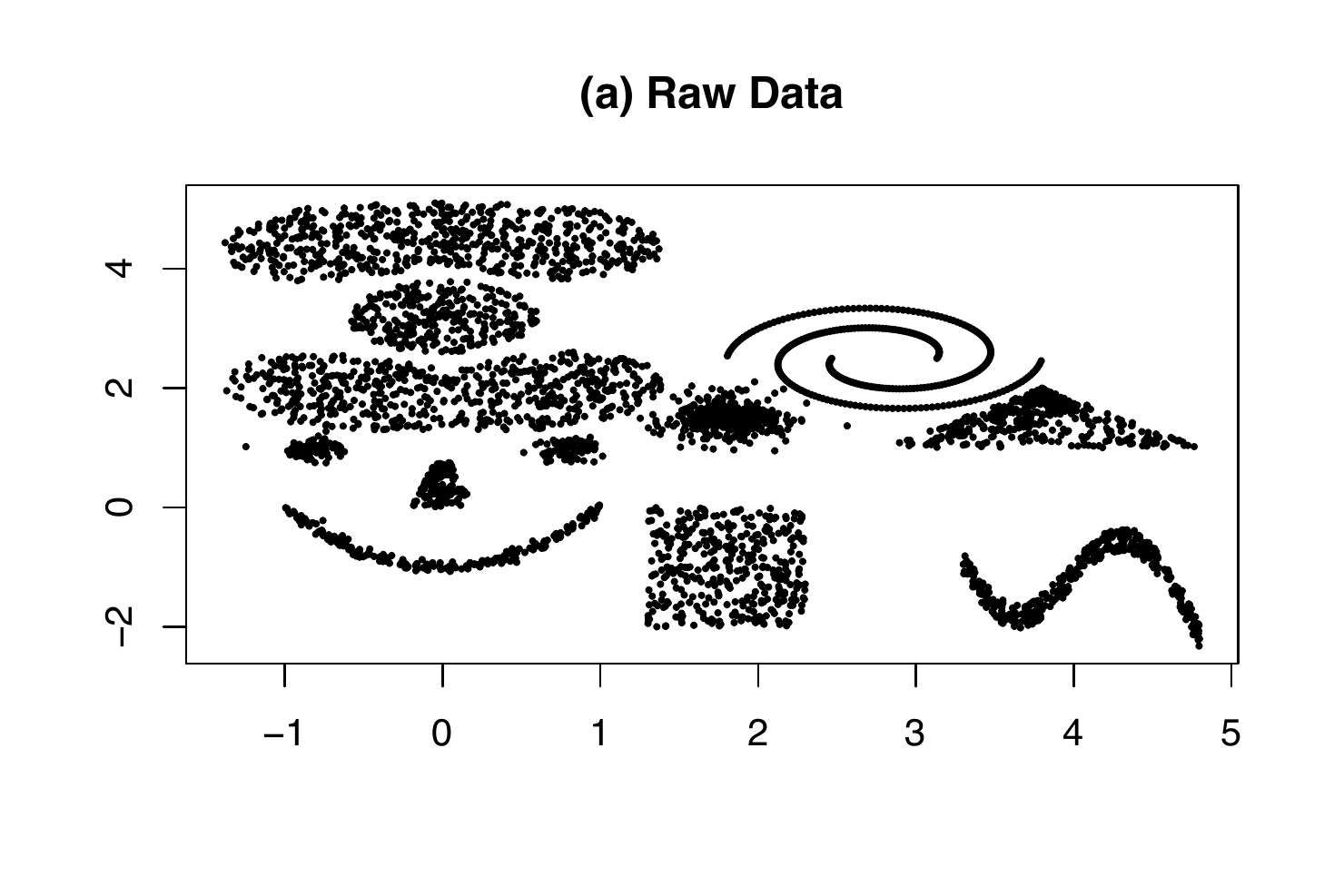}
  \caption*{\small (a) Raw Data}
\endminipage\hspace{0.1cm}%
\minipage{0.3\textwidth}
\centering
  \includegraphics[width=\linewidth]{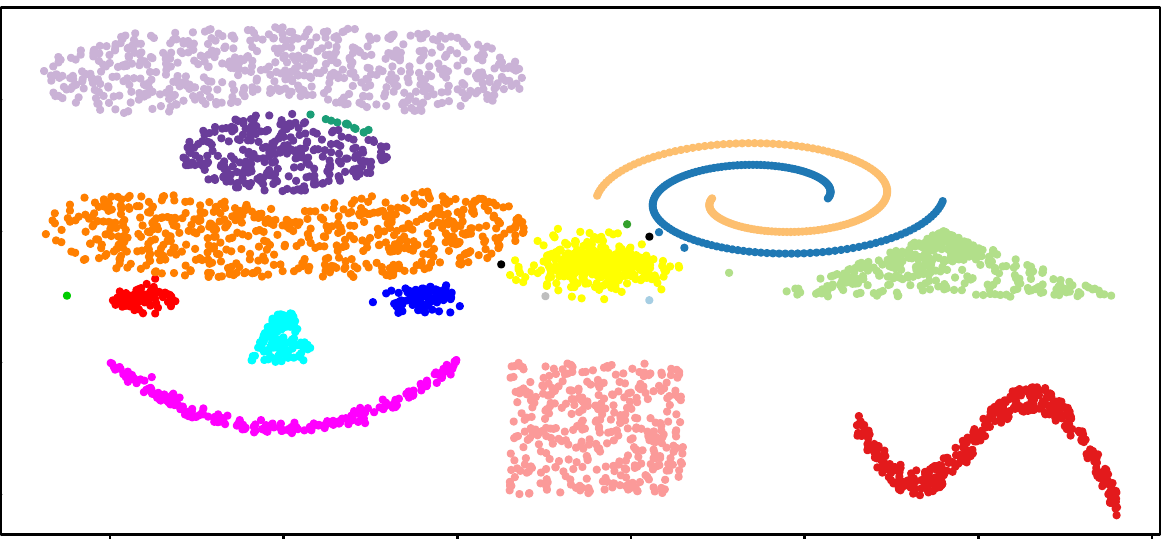}
  \caption*{\small (b) CRAD}
\endminipage\hspace{0.1cm}%
\minipage{0.3\textwidth}%
\centering
  \includegraphics[width=\linewidth]{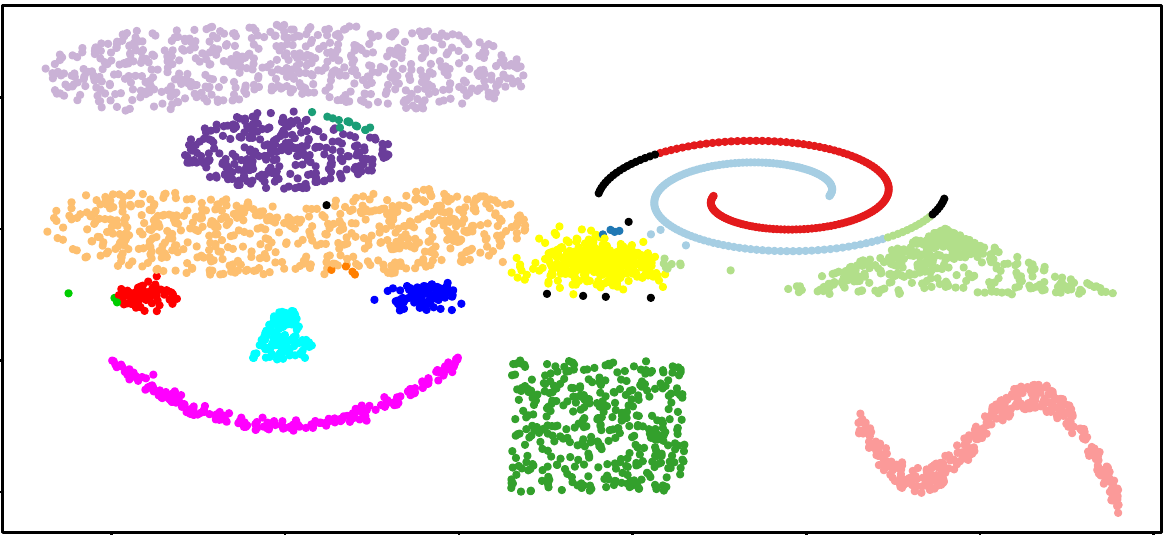}
  \caption*{\small (c) CRAD-DBSCAN}
\endminipage

\vspace*{0.3cm}

\minipage{0.3\textwidth}%
\centering
  \includegraphics[width=\linewidth]{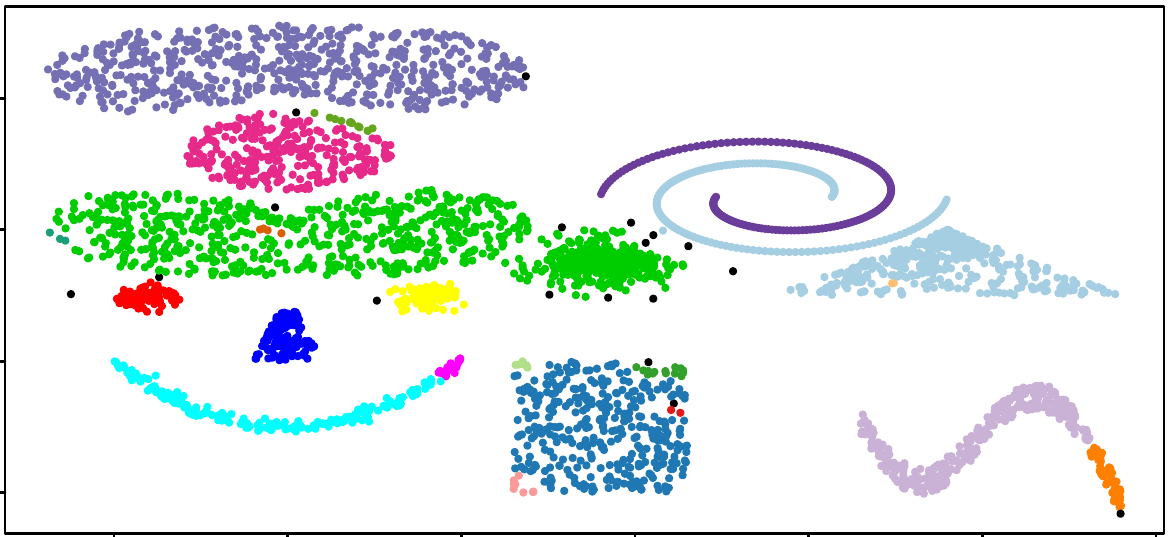}
  \caption*{\small (d) DBCA}
\endminipage\hspace{0.1cm}%
\minipage{0.3\textwidth}%
\centering
  \includegraphics[width=\linewidth]{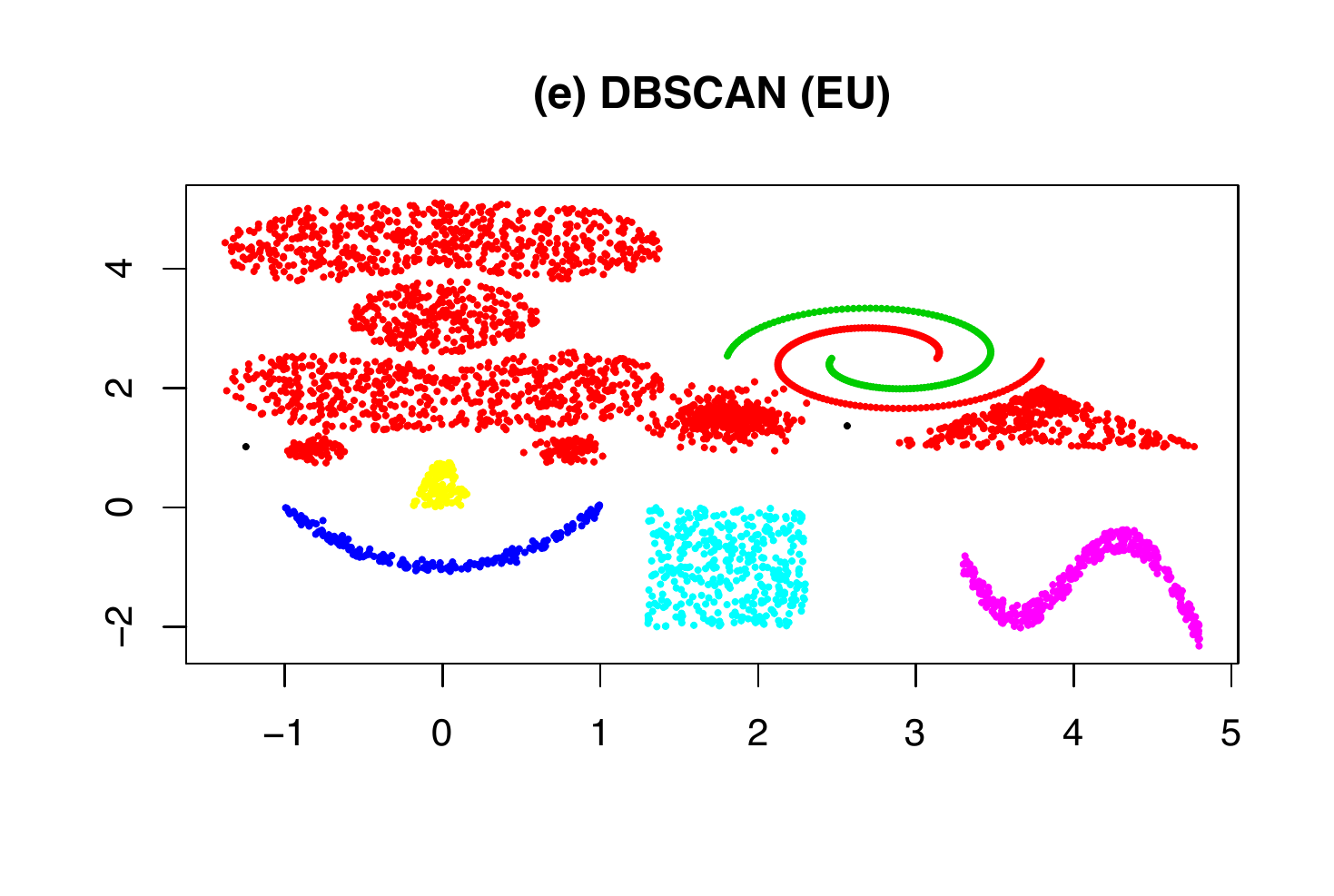}
  \caption*{\small (e) DBSCAN (EU)}
\endminipage\hspace{0.1cm}%
\minipage{0.3\textwidth}%
\centering
  \includegraphics[width=\linewidth]{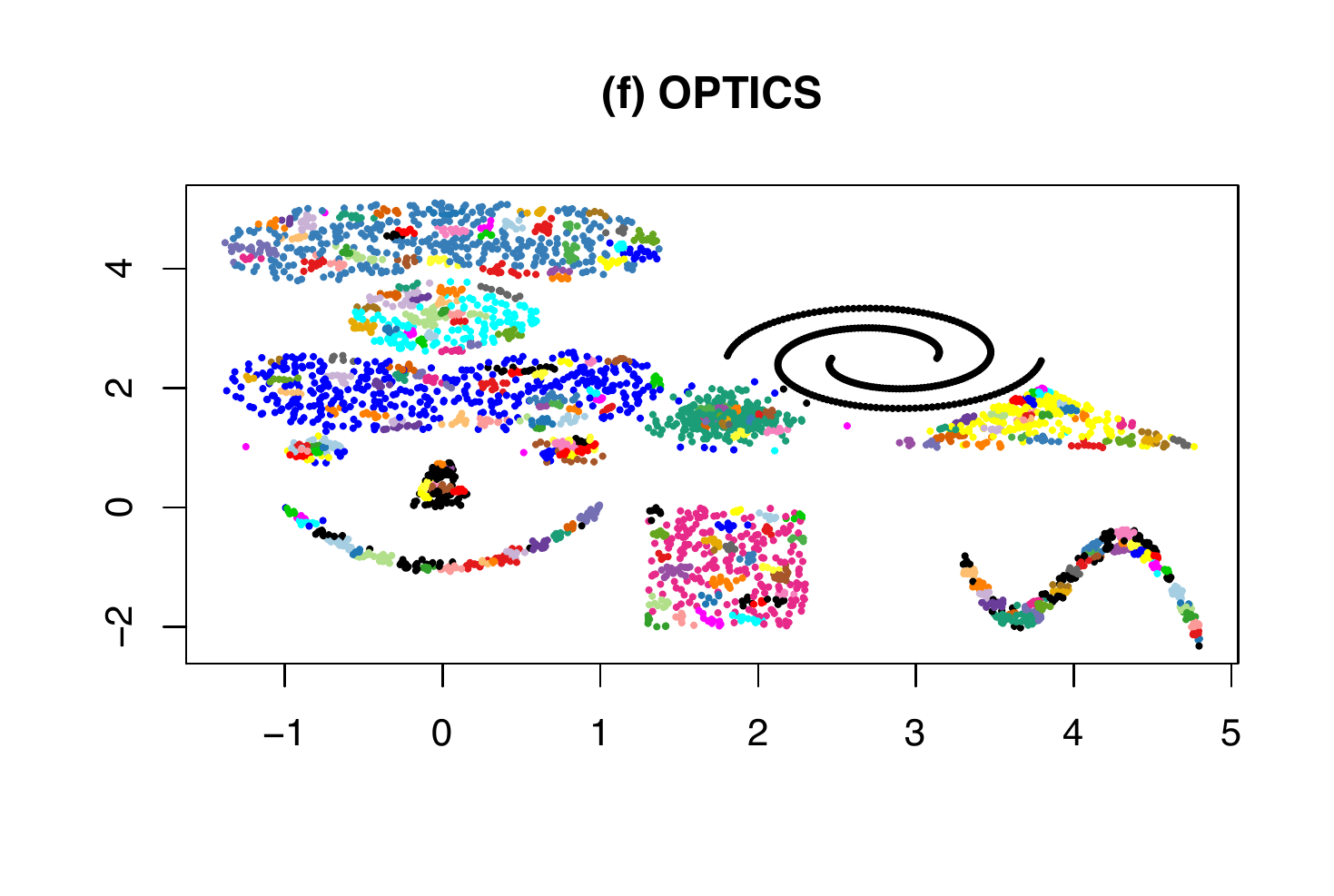}
  \caption*{\small (f) OPTICS}
\endminipage
\caption{Clustering Performance of CRAD, CRAD-DBSCAN, DBCA, DBSCAN (EU), and OPTICS on $S1$.}\label{fig:s1}
\end{figure*}

\begin{figure*}
\centering
\minipage{0.3\textwidth}
\centering
  \includegraphics[width=\linewidth]{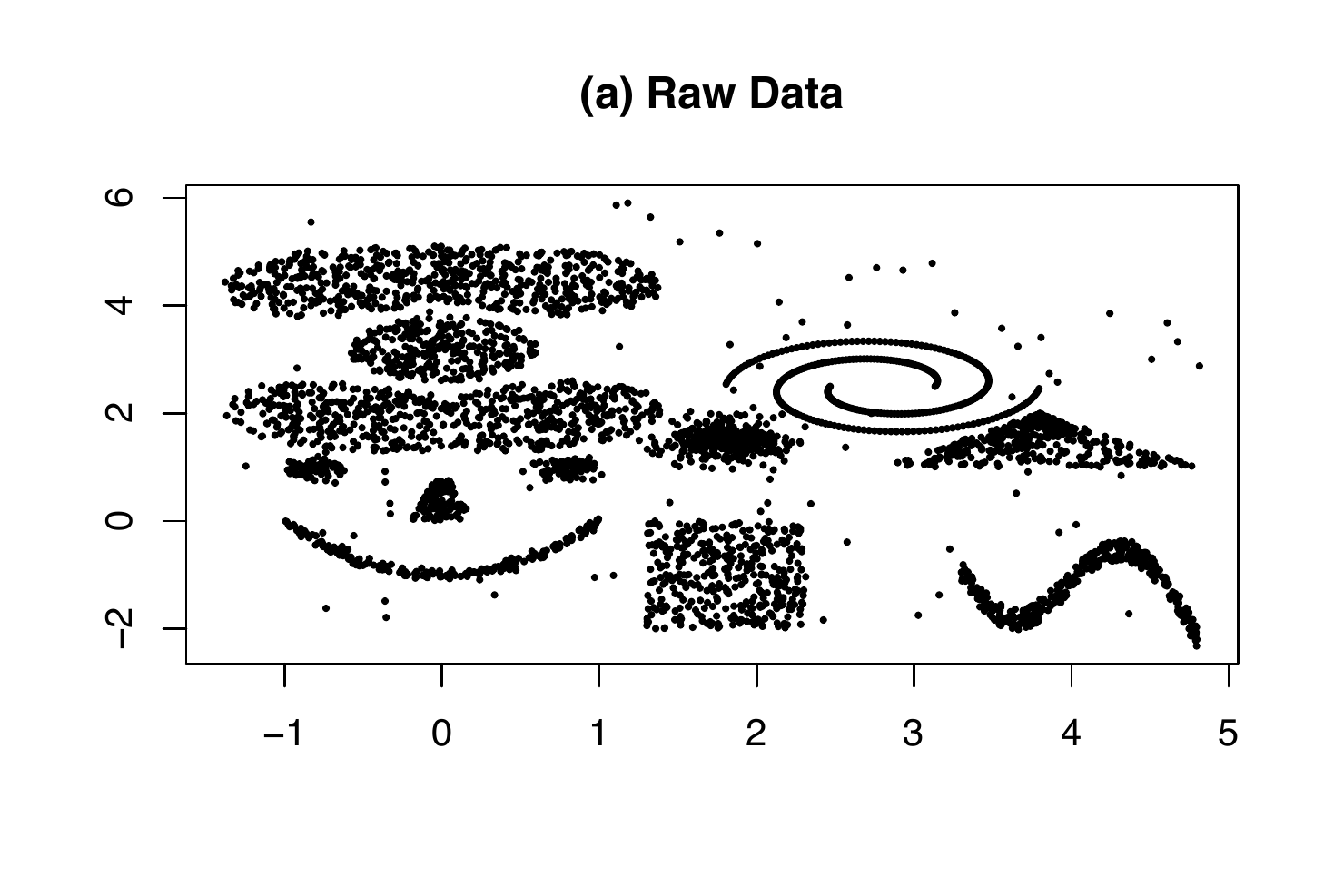}
  \caption*{\small (a) Raw Data}
\endminipage\hspace{0.1cm}%
\minipage{0.3\textwidth}
\centering
  \includegraphics[width=\linewidth]{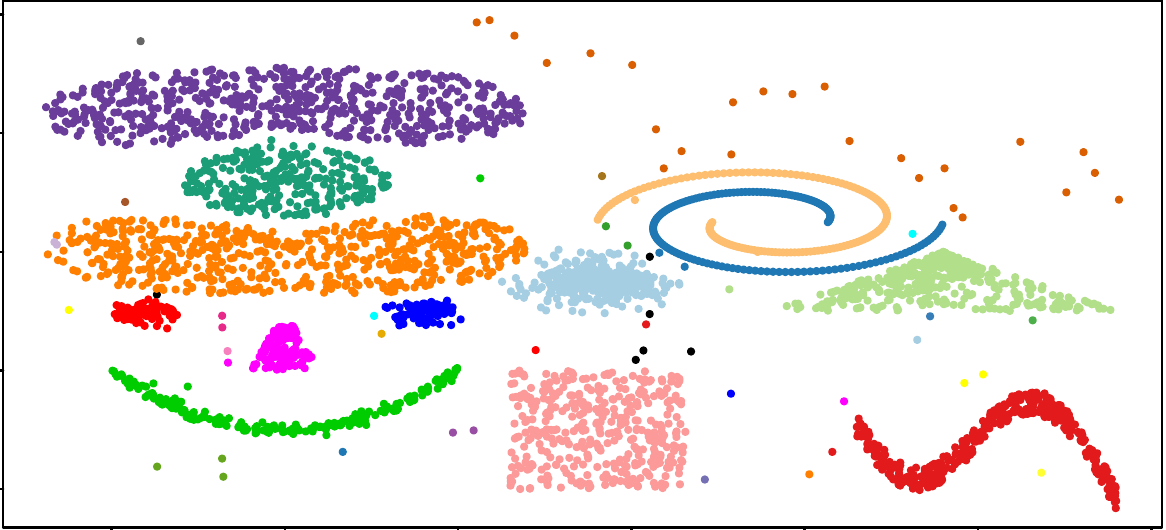}
  \caption*{\small (b) CRAD}
\endminipage\hspace{0.1cm}%
\minipage{0.3\textwidth}%
\centering
  \includegraphics[width=\linewidth]{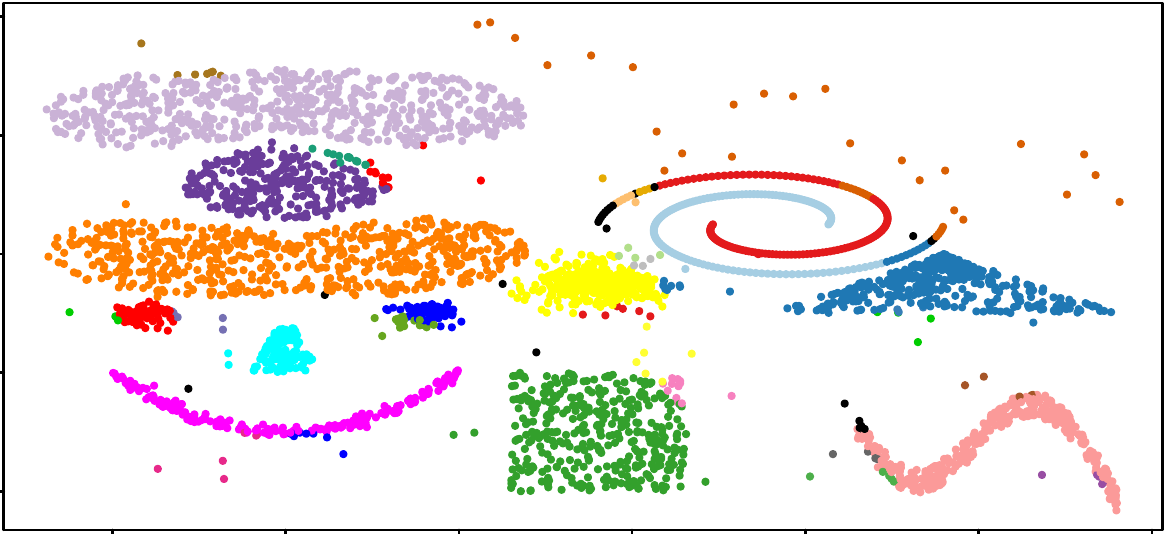}
  \caption*{\small (c) CRAD-DBSCAN}
\endminipage\hspace{0.1cm}%

\vspace*{0.3cm}

\minipage{0.3\textwidth}%
\centering
  \includegraphics[width=\linewidth]{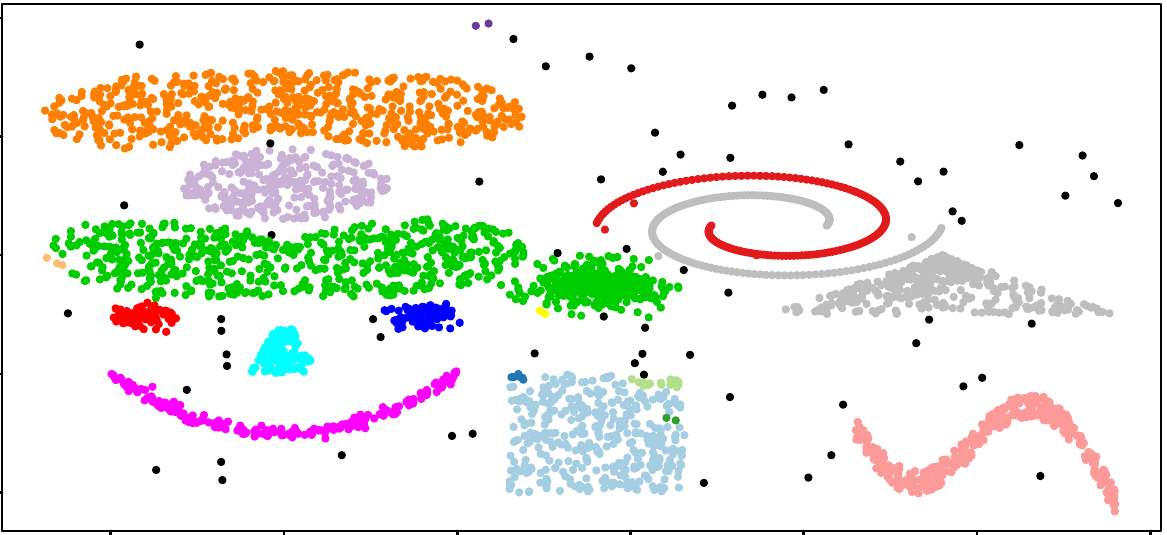}
  \caption*{\small (d) DBCA}
\endminipage\hspace{0.1cm}
\minipage{0.3\textwidth}%
\centering
  \includegraphics[width=\linewidth]{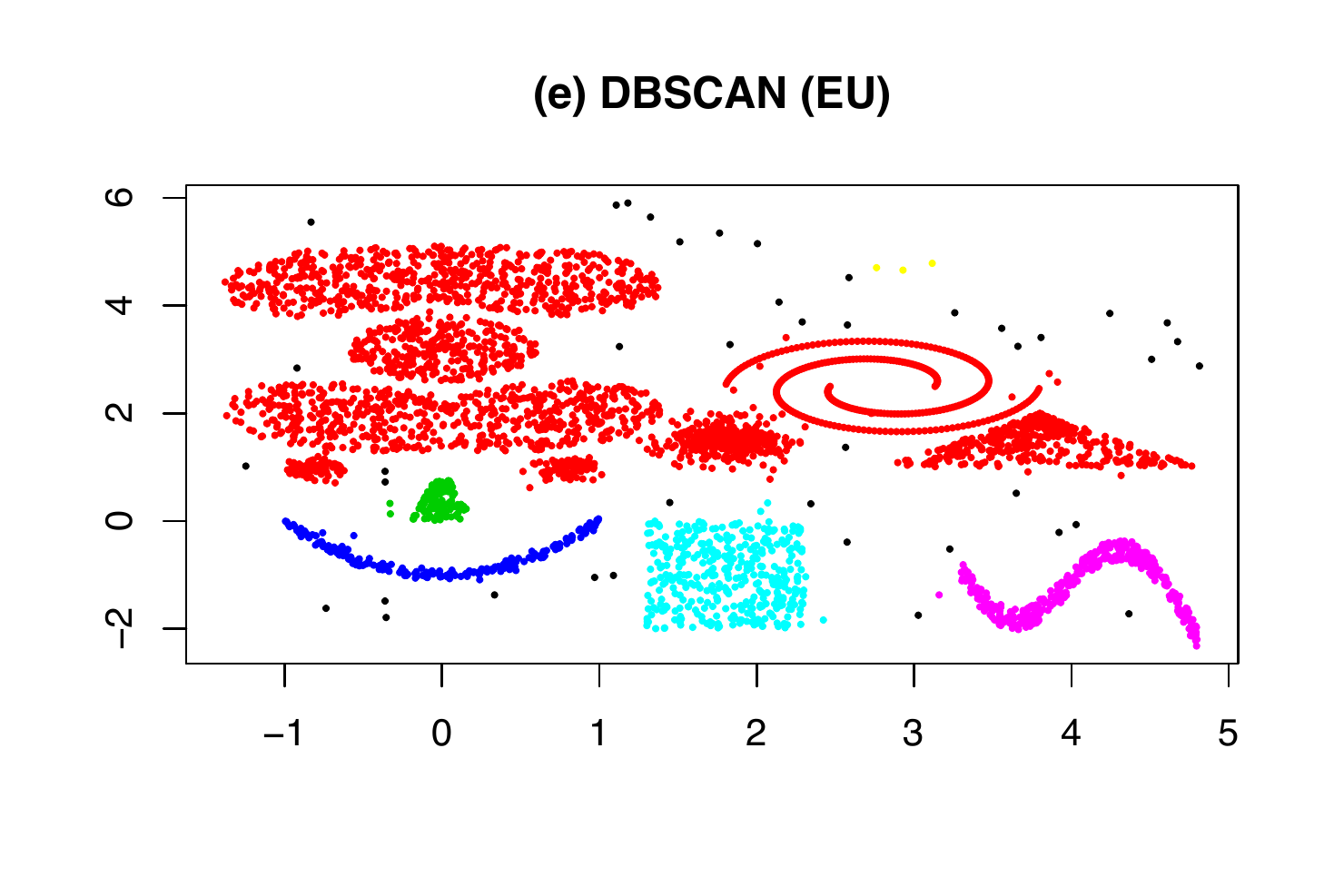}
  \caption*{\small(e) DBSCAN (EU)}
\endminipage\hspace{0.1cm}%
\minipage{0.3\textwidth}%
\centering
  \includegraphics[width=\linewidth]{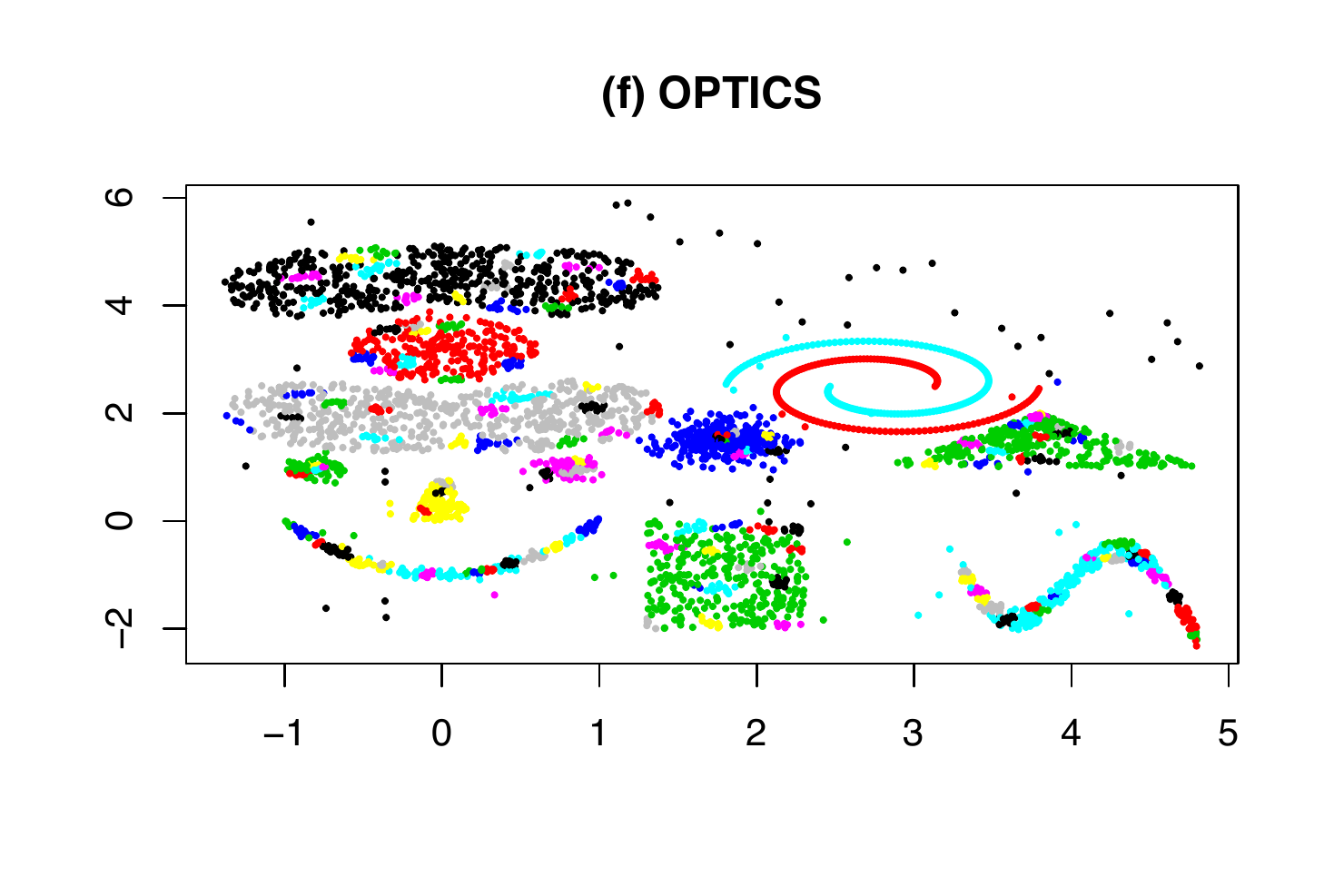}
  \caption*{\small(f) OPTICS}
\endminipage
\caption{Clustering Performance of CRAD, CRAD-DBSCAN, DBCA, DBSCAN (EU), and OPTICS on $S2$.}\label{fig:s2}
\end{figure*}

The evaluation is first conducted on 2 synthetic data sets, $S1$ and $S2$. To visualize the improved effects of our algorithm on DBCA, we extend the data sets in~\cite{jeong2016data} so that they exhibit the challenging properties on which we focus. Specifically, for $S1$ we generate a mixture of clusters from both normal and uniform distributions (with varying density among clusters) by replacing the ``circles" shaped clusters with the ``cassini" cluster structure. In addition, we decrease the distance bewteen clusters, which makes it harder to detect true patterns. All the sample data are from the mlbench~\cite{mlbench:Friedrich2010,Lichman:2013}. The extended dataset $S1$ is shown in Figure~\ref{fig:s1}(a). Lastly, we explore the performance of algorithms under the existence of noises. Dataset $S2$ is generated by adding a number of noises with 2$\%$ noise to signal ratio, shown in Figure~\ref{fig:s2}(a).

As evaluation metric we consider Adjusted Mutual Information (AMI)~\cite{vinh2010information,meilua2007comparing}, which is a robust adjustment of the Mutual Information (MI) score. Given a set $X$ of $n$ observations $(x_1,x_2, \dots ,x_n)$, let us consider two partitions of $X$, namely ${\displaystyle U=\{U_{1},U_{2},\ldots ,U_{R}\}}$ with $R$ clusters, and ${\displaystyle V=\{V_{1},V_{2},\ldots ,V_{C}\}}$ with $C$ clusters. The AMI is defined as follow:
\begin{equation}
	AMI(U,V) = \frac{MI(U,V)- \mathop{\mathbb{E}}(MI(U,V))}{\max\{H(U), H(V)\} - \mathop{\mathbb{E}}\{MI(U, V)\}},
\end{equation}
where
\begin{eqnarray*}
H(U) &=& -\sum\limits_{i=1}^R P(i)\log(P(i)), \\
MI(U,V) &=& \sum\limits_{i=1}^R\sum\limits _{j=1}^C P(i,j)\log\frac{P(i,j)}{P(i)P^{\prime}(j)},
\end{eqnarray*}
$P(i) = |U_i|/N$, $P^{\prime}(j) = |V_j|/n$ and $P(i,j) = |U_i \cap V_j|/n$.
In contrast to MI, the value of AMI between two random clusterings takes on a constant value, especially when the two partitions have a larger number of clusters~\cite{vinh2010information}.

For each clustering algorithm, we search the best achievable clustering result in a wide range of combinations of its parameters. The search range of $Nbin$ in CRAD and CRAD-DBSCAN is in $\{80, 90, \ldots, 700\}$, and $StepSize$ is set as 1. The search range of $\theta$ in DBCA is in $\{0.80, 0.82, \ldots, 1\}$. In DBSCAN (EU), $\epsilon$ is selected from minimum to the half of the maximum value of pairwise dissimilarity in the given dataset. Parameter $MinPts$ for CRAD-DBSCAN, DBCSAN (EU) and OPTICS is selected from $\{2,3,\dots,6\}$, and $\xi \in \{0.01,0.02,\ldots,0.99\}$.%

\begin{figure}[!t]
  \includegraphics[width=\linewidth]{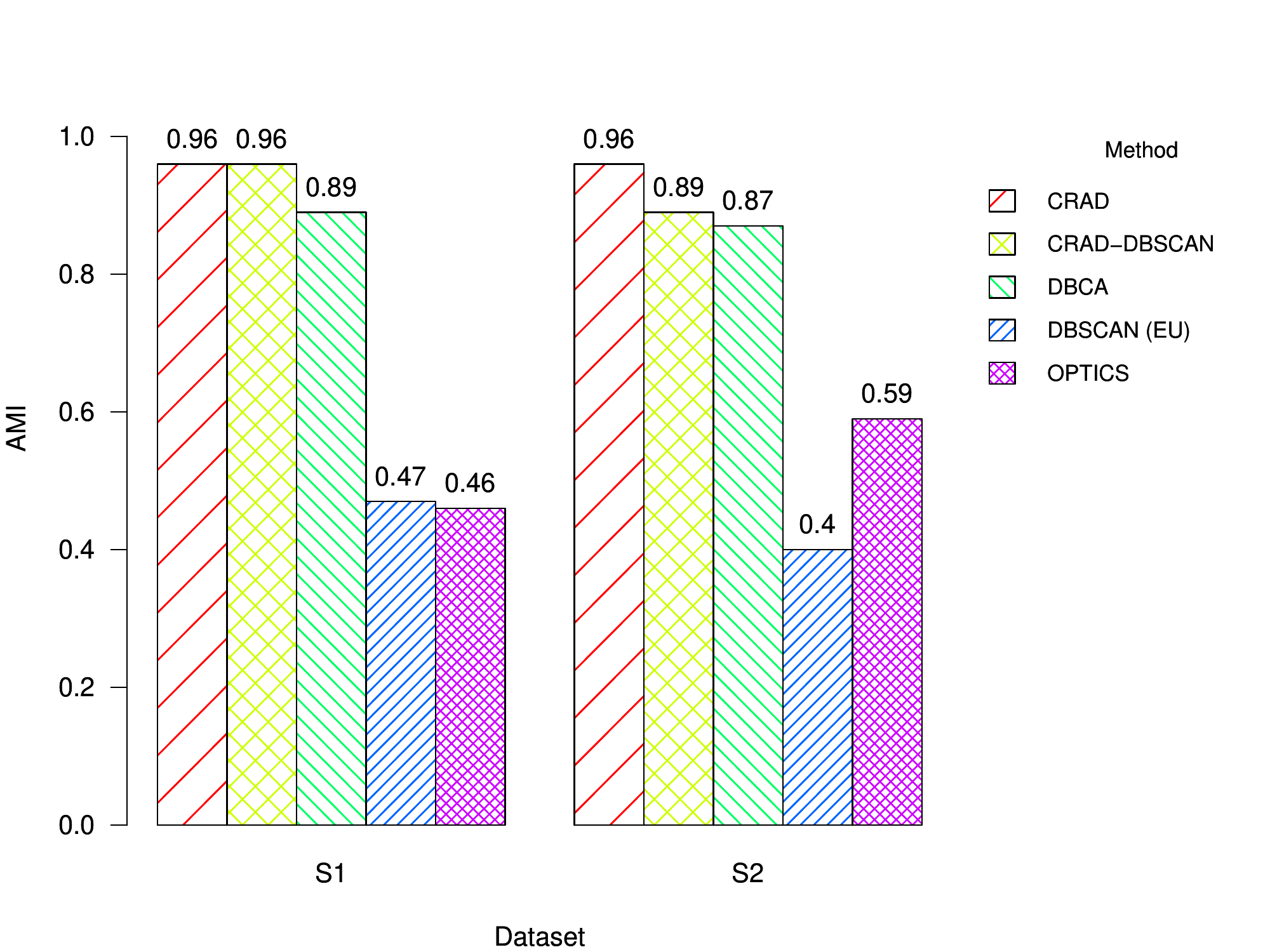}
  \caption{Best AMI of CRAD, CRAD-DBSCAN, DBCA, DBSCAN (EU) and OTPICS on synthetics $S1$ and $S2$. Each AMI score is an average result over 10 trials.}\label{fig:s1_s2}
\end{figure}

The clustering results on $S1$ and $S2$ are shown in Figure~\ref{fig:s1_s2}, where each number is an average result over 10 trails. The clustering results on $S1$ and $S2$ are visualized in Figure~\ref{fig:s1} and Figure~\ref{fig:s2}. For $S1$, we can see that our new CRAD achieves the best clustering performance, with an almost perfect detection result. In addition, CRAD-DBSCAN produces almost the same result as CRAD with a minor misclassification on the boundaries of the ``spiral" cluster (top right). DBSCAN (EU), in contrast, fails to separate most of the clusters, which well demonstrates the competitive performance of our new neighbor searching algorithm. DBCA has a slightly better performance than DBSCAN (EU) but still cannot recognize the ``cassini" cluster (top left) and the ``spiral" cluster (top right) with their nearby clusters. Lastly, OPITCS delivers the poorest performance among the five considered methods. Same conclusion is obtained for $S2$, that is, CRAD and CRAD-DBCSAN outperform all the other competing methods, showing highly competitive performance in detecting intrinsic clusters with varying densities under the existence of noises.

\begin{table*}[!htbp]
\centering
\caption{Clustering Performance on 5 UCI Multivariate Datasets. The winner method on each dataset is highlighted.}\label{table:uciMultiV}
{\renewcommand{\arraystretch}{1}
\begin{tabular}{c| c| c| c| c|c}
\hline
Dataset & \multicolumn{5}{c}{Rand Index}\\
\cline{2-6}
&CRAD & CRAD & DBCA & DBSCAN  & OPTICS\\
& & -DBSCAN & & (EU) & \\
\hline
Banknote & $\mathbf{0.86}$ & $0.79$  & $0.52$ &  $0.81$ & $0.53$ \\

Iris & $0.77$ &     $\mathbf{0.78}$ &    $0.36$  &    $\mathbf{0.78}$ & $0.75$ \\

Blood Transf. & $\mathbf{0.64}$   &  $\mathbf{0.64}$  &    $0.46$  &   $\mathbf{0.64}$ & $0.49$ \\

Occupancy & $\mathbf{0.77}$ &  $0.72$ &  $0.67$  & $0.67$ & $0.66$\\

Seeds & $0.68$  & $0.67$ & $0.33$ & $\mathbf{0.69}$ & $\mathbf{0.69}$ \\
\hline
\end{tabular}}
\end{table*}

 \begin{table*}[!htbp]
 \centering
 \caption{Clustering Performance for the 29 UCR Time Series Datasets. Two ratios, which are over 0.9 and 0.8 in column 	``CRAD / $k$-means" and column ``Empirical CRAD / CRAD", are highlighted, respectively.}\label{table:ucrTs}
 {\renewcommand{\arraystretch}{1.1}
 \begin{tabular}{|c| c| c| c|c|c| c| c |}
 \hline
 Dataset & CRAD/$k$-means & \parbox[t]{2cm}{Empirical CRAD \\/CRAD} & CRAD & $k$-means& Empirical CRAD & \parbox[t]{2cm}{$\#$ of Time Series \\/Length}& $\#$ of Cluster \\
 \hline

50words  & $\mathbf{0.96}$ & $0.35$ & $0.90$ & $0.94$ & $0.31$ & $905/271$& $9$\\
 \hline
 Beef & $\mathbf{1.09}$ & $\mathbf{0.80}$ & $0.76$ &$0.70$ &$0.62$ & $60/471$ & $5$ \\
  \hline
 BirdChicken & $\mathbf{1.09}$ & $\mathbf{1.00}$ & $0.54$ & $0.50$ &  $0.54$ & $40/513$ & $2$ \\
  \hline
 Car &  $\mathbf{1.10}$ & $0.64$ & $0.75$ & $0.68$ & $0.48$ & $120/578$ & $4$ \\
  \hline
 Coffee & $0.88$ & $\mathbf{1.00}$ & $0.74$ & $0.83$ & $0.74$ & $56/287$ & $2$ \\
  \hline
 Cricket-X & $\mathbf{1.01}$ & $\mathbf{0.87}$ & $0.88$ &  $0.87$  & $0.77$ & $780/301$ & $9$ \\
  \hline
 Cricket-Y & $\mathbf{1.02}$ & $\mathbf{0.82}$ & $0.89$ &$ 0.87$ & $ 0.73$ & $780/301$ & $9$\\
  \hline
 Cricket-Z  & $\mathbf{1.02}$ & $\mathbf{0.94}$ & $0.88$ & $0.86$ & $0.83$ & $780/301$ & $9$ \\
  \hline
  ECG200  & $\mathbf{0.97}$  & $\mathbf{0.87}$ &$ 0.61$  & $0.63$ & $0.53$  & $200/97$ & $3$ \\
  \hline
  ECG-FiveDays  & $\mathbf{1.00}$  & $\mathbf{0.82}$  & $0.98$  & $0.98$ & $0.80$ & $5000/141$ &$ 2$ \\
  \hline
  FaceFour &$\mathbf{0.95}$  & $\mathbf{0.98}$ &$0.94$  &$0.99$ & $0.92$ & $112/351$ & $4$ \\
  \hline
  FISH & $\mathbf{1.00}$ & $\mathbf{0.99}$ & $0.84$  & $0.84$  & $0.83$ & $350/464$ & $7$ \\
  \hline
  Gun-Point & $\mathbf{1.03}$ & $0.79$ & $0.77$ & $0.77$ & $0.62$ & $200/151$& $2$ \\
  \hline
  Ham  & $\mathbf{0.96}$ & $\mathbf{1.00}$ & $0.56$ & $0.59$ & $0.56$ & $214/432$ & $2$ \\
  \hline
  Haptics  & $\mathbf{1.10}$  &  $\mathbf{0.88}$ & $0.75$ & $0.68$ & $0.66$  & $463/1093$&$5$ \\
  \hline
  Herring  & $\mathbf{0.96}$ & $\mathbf{0.93}$ & $0.53$ & $0.55$ &$0.49$ & $128/513$& $2$\\
  \hline
  InlineSkate  & $\mathbf{1.09}$ & $0.31$  & $0.82$ & $0.76$ & $0.25$ & $650/1883$ & $7$ \\
  \hline
  Lighting2  & $\mathbf{1.04}$ & $\mathbf{0.93}$ & $0.54$ & $0.52$ & $0.50$ & $121/638$ & $3$ \\
  \hline
  Lighting7  & $\mathbf{1.01}$ & $\mathbf{1.00}$  & $0.82$  & $0.81$ & $0.82$ & $143/320$ & $7$ \\
  \hline
  Meat  & $\mathbf{0.97}$ & $\mathbf{1.00}$ & $0.78$ & $0.80$ &  $0.78$  &$120/449$ &  $3$\\
  \hline
  OSULeaf  & $\mathbf{1.03}$ & $\mathbf{0.87}$ & $0.80$ & $0.77$ & $0.70$ & $442/428$& $6$ \\
  \hline
  Plane  & $\mathbf{0.98}$ & $\mathbf{0.96}$ & $0.95$ & $0.97$ & $0.91$ & $210/145$&$7$ \\
  \hline
  SonyAIBORobotSurface  & $\mathbf{0.94}$ & $\mathbf{1.04}$ & $0.67$ & $0.71$ & $0.70$ & $621/71$&$2$ \\
  \hline
  Synthetic-Control  & $\mathbf{0.96}$ & $\mathbf{1.00}$ & $0.86$  & $0.89$ &$0.86$ & $600/61$ & $6$\\
  \hline
 ToeSegmentation1   & $\mathbf{0.95}$ & $0.77$ & $0.81$ & $0.85$ & $0.62$ & $268/278$&$2$ \\
  \hline
  ToeSegmentation2   & $\mathbf{1.00}$ & $\mathbf{0.95}$ & $0.89$ & $0.89$ & $0.84$ & $166/344$ & $2$ \\
  \hline
  Trace   & $\mathbf{0.99}$ & $\mathbf{1.00}$ & $0.99$ & $1.00$ & $0.99$ & $200/276$ &$4$ \\
  \hline
  Worms & $\mathbf{1.13}$ & $0.66$ & $0.75$ & $0.66$ & $0.49$ & $258/901$&$5$ \\
  \hline
   WormsTwoClass & $\mathbf{0.97}$ & $\mathbf{0.98}$ & $0.54$ & $0.56$ & $0.53$ &$258/901$ & $2$ \\
  \hline

\end{tabular}}
\end{table*}

\subsection{Read-World Multivariate Data}

We now evaluate CRAD on real-world multivariate data, with the same competing methods and experiment settings, as in Section~\ref{section: experimental evaluation}-A. The evaluation is conducted on 5 benchmark multivariate datasets from UCI~\cite{Lichman:2013}: \textit{Banknote Authentication}, \textit{Iris}, \textit{Blood Transfusion}, \textit{Occupancy Detection}, and \textit{Seeds}. A brief description of each dataset is shown as follows.

\begin{itemize}[leftmargin=*]
\item  \textit{Banknote Authentication}: contains 762 and 610 observations for each of two classes of banknotes, respectively. Each observation has 4 attributes, which are extracted features from the image of the banknote-like specimen.

\item \textit{Iris}: contains 3 classes of 50 observations each, where each class refers to a type of iris plant. The number of attributes for each observation is 4.

\item \textit{Blood Transfusion}: contains 570 and 178 observations for each of two classes of people, respectively. Each class represents whether he/she donated blood in March 2007. The number of attributes for each observation is 4.

\item \textit{Occupancy Detection}: contains 7703 and 2049 observations for each of two classes of office rooms, respectively. Each class indicates whether the office room is occupied. Each observation has 5 attributes, which are temperature, humidity, light, 
CO2 and humidity ratio in the room.

\item \textit{Seeds}: contains 3 classes of 70 observations each, where each class refers one kind of wheat. Each observation has 7 attributes, which are geometric descriptions of the wheat kernel.

\end{itemize}

As Table~\ref{table:uciMultiV} indicates, for all datasets, except \textit{Seeds}, CRAD and CRAD-DBSCAN rank \nth{1}/\nth{2} among all the methods. In particular, CARD and CRAD-DBSCAN significantly outperform DBCA and OPTICS for \textit{Banknote Authentication}, \textit{Iris}, \textit{Blood Transfusion}, and \textit{Occupancy Detection}. Furthermore, CRAD outperforms DBSCAN (EU) for \textit{Banknote Authentication} and \textit{Occupancy Detection} and delivers a comparable performance for \textit{Iris} and \textit{Blood Transfusion}. For \textit{Seeds}, CRAD and CRAD-DBSCAN slightly underperform, comparing to DBSCAN (EU) and OPTICS, but still significantly outperform DBCA.

\subsection{Real-World Time Series Data}
We now evaluate the utility of CRAD for time series clustering. Time series data usually contain noises, dropouts, or extraneous data, existence of which can greatly limit the accuracy of clustering \cite{ye2009time,mueen2011logical,hartmann2010prototype}. Thus, we apply a time-series based feature-extraction technique, named U-Shapelets~\cite{zakaria2012clustering} to filter out noises in data in the first place. The idea of the U-Shapelets is to search for small subsequences of a few time series, named U-Shapelets, that best represent the entire time series data and then to use those subsequences as features. Since the number of extracted U-Shapelets is small (usually $<10$), dimension of time series data is highly reduced.

Based on the extracted U-Shapelets, we evaluate our CRAD algorithm with respect to the ``U-Shapelets + $k$-means" methodology, where $k$ denotes the true number of clusters~\cite{zakaria2012clustering}.
The choice for the study settings is addressed as follows.
\begin{enumerate}[leftmargin=*]
\item First, we select ``U-Shapelets + $k$-means" as the competing clustering method, because as demonstrated by~\cite{zakaria2012clustering}, it is the winner method over other clustering methods with the state-of-the-art feature-extraction techniques.
\item Second, the density-based clustering algorithms such as DBSCAN, DBCA, and OPTICS are not included as competing methods, as none of these methods are designed for time series clustering. Furthermore, we focus on the following practical problems in this section:
\begin{enumerate}
\item
Without any a-priori knowledge on a number of clusters, can our CRAD algorithm detect the true number of clusters and the correct partitions?
\item
Without knowing the true parameters, $Nbin$ and $StepSize$, can our parameter selection procedure, from Section~\ref{section: determine para}, assist the clustering method to achieve a satisfactory clustering performance?
\end{enumerate}
\end{enumerate}

The evaluation is conducted on 29 benchmark datasets from the UCR time series archive
~\cite{UCRArchive}, in terms of RI~\cite{rand1971objective} and is consistent with the evaluation of ``U-Shapelets + $k$-means" in~\cite{zakaria2012clustering}. The datasets include various domains, i.e., from finance to neuroscience to geology (see Table~\ref{table:ucrTs}). The column ``CRAD" and ``$k$-means" denote the best achievable RI by searching clustering results over a wide range of parameters and possible combinations of U-Shapelets features. As shown in~\cite{zakaria2012clustering}, 1 or 2 U-Shapelets are sufficient to achieve the best clustering result in most cases. Hence, the upper limit on a number of U-Shapelets is set to 2.
The column ``Empirical CRAD" is the RI achieved by CRAD, using the new parameter selection procedure in Section~\ref{section: determine para}. Compared with the best achievable ``CRAD", the ``Empirical CRAD" is more important since we will not know the ground truth in real-data clustering and thus being able to select the right parameter is critical in achieving a good clustering result. Two ratio indicators, that is ``CRAD / $k$-means" and ``Empirical CRAD / CRAD", are presented to simplify the comparison among methods.

For the ratio ``CRAD / $k$-means", 28 (out of 29) datasets are over 0.9, among which 16 datasets deliver a ratio of more than 1, indicating the competitive performance of our CRAD algorithm. Note, our benchmark method ``U-Shapelets + $k$-means" has
a critical advantage of knowing the true number of clusters in datasets, thus operating with more information.
Despite this, the new CRAD algorithm still delivers a quite close performance with the benchmark method and even outperforms it in half of the datasets. For the ratio ``Empirical CRAD / CRAD", 23 (out of 29) datasets are over 0.8, among which 16 datasets yield a ratio more than 0.9, and 13 datasets deliver a ratio more than 0.95. These findings  indicate a high practical utility of CRAD in the real-world time-series clustering, which is typically performed without any prior information on the true number of clusters and cluster density.  



\medskip
Finally, we assess performance of CRAD with respect to other density-based clustering algorithms. The competing methods are DBCA and DBSCAN.
All the methods are performed on U-Shapelets extracted from the data, following the framework of clustering time series in~\cite{zakaria2012clustering}, i.e., ``U-Shapelets + a clustering method". The time series data are selected from Table~\ref{table:ucrTs}. Each dataset contains two versions: a raw dataset and a noisy dataset which is obtained by adding a random noise $N(0,0.2)$ on each observation of time series in the raw dataset.

\begin{table}[!htbp]
\footnotesize
\centering
\caption{Clustering Performance of CRAD, DBCA and DBSCAN on 5 UCR Time Series Datasets. The winner method on each dataset is highlighted.}\label{table:ucrTs_2}
{\renewcommand{\arraystretch}{1}
\begin{tabular}{c| c| c| c}
\hline
Dataset & \multicolumn{3}{c}{Rand Index}\\
\cline{2-4}
&CRAD & DBCA & DBSCAN \\
\hline
Coffee & $0.74$ & $0.59$  & $\mathbf{0.75}$ \\
Noisy Coffee & $\mathbf{0.65}$ & $0.58$  & $0.55$ \\
\hline
\hline
FaceFour & $0.94$ &     $ 0.91$ &    $\mathbf{0.98}$ \\
Noisy FaceFour & $\mathbf{0.92}$ &     $ 0.86$ &    $0.85$ \\
\hline
\hline

SonyAIBO & $0.67$   &  $\mathbf{0.68}$  &    $0.53$ \\
Noisy SonyAIBO & $\mathbf{0.67}$   &  $0.56$  &    $0.52$ \\
\hline
\hline
ToeSegm1 & $\mathbf{0.81}$ &  $0.71$ &  $0.72$ \\
Noisy ToeSegm1 & $\mathbf{0.80}$ &  $0.68$ &  $0.68$ \\
\hline
\hline
Trace & $0.99$  & $0.99$ & $\mathbf{1.00}$ \\
Noisy Trace & $\mathbf{0.95}$  & $0.93$ & $0.89$ \\
\hline
\end{tabular}}
\end{table}

As Table~\ref{table:ucrTs_2} indicates, on the raw datasets CRAD outperforms DBCA and DBSCAN for \textit{ToeSegm1} and delivers a comparable performance for \textit{Coffee}, \textit{FaceFour}, \textit{SonyAIBO}, and \textit{Trace}. However, under noised scenarios, CRAD outperforms DBCA and DBSCAN on all the five considered datasets. These findings are consistent with conclusions in previous sections, that is, our new CRAD algorithm delivers a more competitive performance for data that contain noise, outliers and of varying densities.

\section{Conclusion}\label{section: conclusion}

We propose a new robust data depth based clustering algorithm CRAD with a locally-defined neighbor searching function. Besides robustness to outliers, we show that the new CRAD algorithm is highly competitive in detecting clusters with varying densities, compared with the existing algorithms such as DBSCAN, OPTICS and DBCA. 
Furthermore, the performance of DBSCAN is shown to be effectively improved, by replacing its original neighbor searching function with the new locally tuned neighbor searching algorithm.
In addition, we propose a new effective parameter selection procedure, to select the optimal underlying parameter in the real-world clustering, when the ground truth is unavailable. In the future, we plan to investigate the utility of other data depth functions as dissimilarity measures and extend the CRAD idea to functional data clustering. 



\section*{Acknowledgment}
The authors would like to thank Karianne Bergen and Cuneyt Akcora for stimulating discussion.
This research is supported by NSF under contract number IIS-1633331.



\bibliographystyle{IEEEtran}
%

\bibliography{KDDbiblio}

%
%

\end{document}